\documentclass[aps,prc,amsmath,amssymb,preprint,floats,nofootinbib,superscriptaddress,color,groupedaddres]{revtex4}
\usepackage[dvips]{graphicx}
\usepackage[dvips]{color}
\usepackage{dcolumn}
\usepackage{bm}

\newcommand{\n}[1]{\ensuremath{|\mathbf{#1}|}}
\newcommand{\ve}[1]{\ensuremath{\mathbf{#1}}}
\newcommand{\bea}{\begin{eqnarray}}
\newcommand{\eea}{\end{eqnarray}}
\newcommand{\be}{\begin{equation}}
\newcommand{\ee}{\end{equation}}
\newcommand{\ba}{\begin{eqnarray}}
\newcommand{\ea}{\end{eqnarray}}

\begin{document}
%\draft
%%%%%%%%%%%%%%%%%%%%%%%%%%%%%%%%%%%%%%%%%%%%%%%%%%%%%%%%%%%%%%%%%%%%%%

\title{Scaling Function, Spectral Function and Nucleon Momentum Distribution in Nuclei}

%%%%%%%%%%%%%%%%%%%%%%%%%%%%%%%%%%%%%%%%%%%%%%%%%%%%%%%%%%%%%%%%%%%%%%
\author{A.~N.~Antonov}
\affiliation{Institute for Nuclear Research and Nuclear Energy, Bulgarian Academy of Sciences,
Sofia 1784, Bulgaria}

\author{M.~V.~Ivanov}
\affiliation{Institute for Nuclear Research and Nuclear Energy, Bulgarian Academy of Sciences,
Sofia 1784, Bulgaria}
\affiliation{Grupo de F\'{i}sica Nuclear, Departamento de F\'{i}sica At\'omica,
Molecular y Nuclear, Facultad de Ciencias  F\'{i}sicas,
Universidad Complutense de Madrid, Madrid E-28040, Spain}

\author{J.~A. Caballero}
\affiliation{Departamento de F\'{i}sica At\'omica, Molecular y Nuclear, Universidad de Sevilla,
41080 Sevilla, SPAIN}

\author{M.~B.~Barbaro}
\affiliation{Dipartimento di Fisica Teorica, Universit\`{a} di
Torino and INFN,\\
Sezione di Torino, Via P. Giuria 1, 10125 Torino, Italy}

\author{J.~M.~Udias}
\affiliation{Grupo de F\'{i}sica Nuclear, Departamento de F\'{i}sica At\'omica,
Molecular y Nuclear, Facultad de Ciencias  F\'{i}sicas,
Universidad Complutense de Madrid, Madrid E-28040, Spain}

\author{E.~Moya de Guerra}
\affiliation{Grupo de F\'{i}sica Nuclear, Departamento de F\'{i}sica At\'omica,
Molecular y Nuclear, Facultad de Ciencias  F\'{i}sicas,
Universidad Complutense de Madrid, Madrid E-28040, Spain}

\author{T.~W.~Donnelly}
\affiliation{Center for Theoretical Physics, Laboratory for Nuclear Science and
Department of Physics,\\ Massachusetts Institute of Technology, Cambridge, MA 02139, USA}

\date{\today}

%%%%%%%%%%%%%%%%%%%%%%%%%%%%%%%%%%%%%%%%%%%%%%%%%%%%%%%%%%%%%%%%%%%%%%%

\begin{abstract}

The link between the scaling function extracted from the analysis of
$(e,e')$ cross sections and the spectral function/momentum
distribution in nuclei is revisited. Several descriptions of the
spectral function based on the independent particle model are
employed, together with the inclusion of nucleon correlations, and
effects of the energy dependence arising from the width of the hole
states are investigated. Although some of these approaches provide rough
overall agreement with data, they are not found to be capable of
reproducing one of the distinctive features of the experimental
scaling function, namely its asymmetry. However, the addition of
final-state interactions, incorporated in the present study using
either relativistic mean field theory or via a complex optical
potential, does lead to asymmetric scaling functions in accordance
with data. The present analysis seems to indicate that final-state
interactions constitute an essential ingredient and are required to
provide a proper description of the experimental scaling function.

\end{abstract}

%%%%%%%%%%%%%%%%%%%%%%%%%%%%%%%%%%%%%%%%%%%%%%%%%%%%%%%%%%%%%%%%%%%%%%%%
%%%%%%%%%%%%%%%%%%%%%%%%%%%%%%%%%%%%%%%%%%%%%%%%%%%%%%%%%%%%%%%%%%%%%%%%%

\maketitle

\section{Introduction}\label{sect1}

Investigations of inclusive quasielastic (QE) electron-nucleus
scattering make it possible to obtain information about one of the
main characteristics of nuclear structure, namely, the spectral
function $S(p,{\cal E})$ and its integral, the nucleon momentum
distribution $n(p)$ in
nuclei~\cite{01Day:1990mf,02Ciofi87,03Ciofi92}. This provides
insights into the validity of the mean-field approximation (MFA) and
the role of the nucleon-nucleon (NN) correlations, as well as into the
effects of Final-State Interactions (FSI) for inclusive electroweak
processes. Using the shell model, it is possible in principle to
obtain the contributions of different shells to $S(p,{\cal E})$ and
the momentum distribution for each single-particle state. However,
due to the residual interactions, the hole states are not
eigenstates of the residual nucleus but are mixtures of several
single-particle states. This leads to the spreading of the shell
structure and only a careful study of the momentum dependence of
$S(p,{\cal E})$ can separate the contributions from different shells
(see, {\it e.g.}~\cite{04ant02}). Such analyses have been carried
out for few-body systems, complex nuclei and nuclear matter, focused
mainly on the existence of high-momentum components of the nucleon
momentum distribution due to NN correlation
effects~\cite{02Ciofi87,03Ciofi92,05Ji89,06Ciofi09,04ant02,07ant02a}.
Since it is impossible within the MFA to describe simultaneously the
density and momentum distributions in nuclei
\cite{08Bohigas80,09Jaminon85,10MoyadeGuerra91,11ant01,04ant02,07ant02a,Barbaro:2009iv},
a consistent analysis of the role of the NN correlations is required
using theoretical methods that go beyond the MFA to obtain a
successful description of the relevant experiments. The present
study uses the results found from studies of $y$-scaling
(\cite{01Day:1990mf,02Ciofi87,03Ciofi92,12West91,13Sick80,14scaling88})
and superscaling (based on $\psi$-scaling variable, {\it e.g.}
\cite{14scaling88,15ant11,16DS199,17DS299,18MDS02,19ant04,20ant05,21ant06,22ant07})
obtained from  analyses of inclusive electron scattering data. The
latter consists in constructing a ``superscaling function''
$f(\psi)$ obtained by removing the single-nucleon content from the
double differential cross section and plotting it versus a scaling
variable $\psi(q,\omega)$. Scaling of the first kind of the scaling
function ({\it i.e.,} no explicit $q$-dependence of $f(\psi)$) can
be seen at excitation energies below the QE peak. Scaling of second
kind ({\it i.e.,} no dependence of $f(\psi)$ on the mass number)
turns out to be excellent in the same region. When scaling of both
first and second types occur, one says that superscaling takes
place. It was pointed out (see, {\it
e.g.}~\cite{17DS299,19ant04,20ant05,21ant06,22ant07}) that the
physical reason for the superscaling is the specific high-momentum
tail of $n(p)$ which arises due to NN correlations and is similar
for all nuclei. As was pointed out in~\cite{23ant10}, however, a
direct connection between the scaling function extracted from the
analysis of the cross section data, and the spectral function only
exists when one makes very restrictive approximations. Along this
line, caution should be kept in mind for the conclusions reached
about the momentum distribution, because a close relationship
between the latter and the scaling function also only emerges after
some approximations are made. In particular, these are linked to the
integration limits involved and the behavior of the spectral
function~\cite{01Day:1990mf}. In~\cite{23ant10} the analysis applied
in the past to the scaling region (that is, negative values of the
scaling variable $y$) was extended to positive $y$, leading to
results that differ from those based solely on the scaling region
and providing new insights into the issue of how the energy and
momentum are distributed in the spectral function.

Under certain approximations ({\it e.g.,} see~\cite{23ant10} and
references therein), in the case of plane waves the ($e,e'N$)
differential cross section factorizes in the form: \be
\left[\frac{d\sigma}{d\epsilon'd\Omega'dp_Nd\Omega_N}
\right]_{(e,e'N)}^{PWIA}= K\sigma^{eN}(q,\omega;p,{\cal
E},\phi_N)S(p,{\cal E})\,,\label{PWIA} \ee where $\sigma^{eN}$ is
the electron-nucleon cross section for a moving off-shell nucleon,
$S(p,{\cal E})$ is the spectral function that gives the probability
to find a nucleon of certain momentum and energy in the nucleus (see
{\it e.g.}~\cite{24Frullani84,25Bof96,26Kel96}) and $K$ is a
kinematical factor~\cite{27Ras89}. In Eq.~(\ref{PWIA}) $p$ is the
missing momentum and ${\cal E}$ is the excitation energy that is
essentially the missing energy minus the separation energy. Further
assumptions are necessary~\cite{23ant10} to show how the scaling
function $F(q,\omega )$ emerges from the Plane-Wave Impulse
Approximation (PWIA), namely, the spectral function is assumed to be
isospin independent and $\sigma ^{eN}$ is assumed to have a very
mild dependence on $p$ and $\cal E$. The scaling function can be
expressed in terms of the differential cross section for inclusive
QE ($e,e'$) processes: \be F(q,\omega)\cong
\dfrac{\left[d\sigma/d\epsilon'd\Omega'\right]_{(e,e')}}{\overline{\sigma}^{eN}
(q,\omega;p=|y|,{\cal E}=0)}\,, \label{scaling} \ee where
$\overline{\sigma}^{eN}$ represents the azimuthal angle-averaged
single-nucleon cross section that also incorporates the kinematical
factor $K$:
\[
\overline{\sigma}^{eN}\equiv
K\sum_{i=1}^A\int d\phi_{N_i}\dfrac{\sigma^{eN_i}}{2\pi} .
\]
Note that in Eq.~(\ref{scaling}) $\overline{\sigma}^{eN}$ is taken
at $p=|y|$, where the magnitude of the scaling variable $y$ is the
smallest value of the missing momentum $p$ that can occur in the
process of electron-nucleus scattering for the smallest possible
value of the excitation energy (${\cal E}=0$), {\it i.e.,} at the
smallest value of the missing energy. Accordingly, in the PWIA the
scaling function $F(q,\omega )$ from Eq.~(\ref{scaling}) may be
expressed in terms of the spectral function: \be F(q,\omega)=
2\pi\int\!\!\!\int_{\Sigma(q,\omega)}p\,dp\, d{\cal E}\,S(p,{\cal
E}) \, , \label{scaling_function} \ee where ${\Sigma(q,\omega)}$
represents the kinematically allowed region (for details, see {\it
e.g.}~\cite{23ant10}). Only in the case when it is possible to
extend the region ${\Sigma(q,\omega)}$ to infinity in the excitation
energy plane ({\it i.e.,} at ${\cal E}_{\max}\rightarrow \infty$),
would the scaling function be directly linked to the momentum
distribution of the nuclear system: \be n(p)= \int_0^\infty d{\cal
E} S(p,{\cal E}).\label{nk} \ee

It was shown from the analyses of the inclusive electron-nucleus
scattering that at high values of the momentum transfer the
extracted scaling function $F_\text{exp} (q,\omega )$ becomes a
function only of the scaling variable $y$, and not of
$q$~\cite{01Day:1990mf,16DS199,17DS299,18MDS02}. It was emphasized
in~\cite{23ant10} that Eq.~(\ref{scaling_function}) does not apply
to $F_\text{exp} (q,\omega )$ because of ingredients not included in
the PWIA, such as final-state interactions, meson-exchange currents
(MEC), rescattering processes, {\it etc.}

Using the Relativistic Fermi Gas model (RFG) as a guide, the
separate analysis of longitudinal ($L$) and transverse ($T$)
($e,e'$) data made it possible to introduce three ``universal''
experimental dimensionless superscaling functions: \be
f_\text{exp}(q,\omega)\equiv k_A
F_\text{exp}(q,\omega);~f^{L(T)}_\text{exp}(q,\omega) \equiv k_A
F^{L(T)}_\text{exp}(q,\omega)\, ,\label{sscalingf} \ee $k_A$ being a
phenomenological characteristic momentum scale for the specific
nucleus being studied; this is equal to the Fermi momentum $k_F$ in
the case of the RFG. In the present work we consider only the
longitudinal scaling function and henceforth drop the subscript
``L'' for simplicity. Note that the effects of FSI and relativity on
this function are often important and, as emphasized
in~\cite{23ant10}, any conclusion about the momentum distribution
based on Eq.~(\ref{scaling_function}) should be made with caution.

In the present work we study in more detail the relationship between
the spectral function $S(p,{\cal E})$ and the scaling function
$F(q,y)$. Our aim is to extract more information about the spectral
function from the experimentally known scaling function, keeping in
mind the restrictions of the PWIA. We take into account the effects
of FSI and some other peculiarities of electron-nucleus scattering.
We make an attempt to construct a spectral function that corresponds
to the experimentally established scaling function following a
series of steps on increasing complexity. Firstly, we construct
$S(p,{\cal E})$ within and beyond the Independent Particle Shell
Model (IPSM). Secondly, we take into account FSI by computing the
inclusive electron-nucleus cross section using the Dirac optical
potential. We incorporate these results in the determination of the
spectral function and consequently the superscaling function. In all
steps we relate the results obtained for the scaling function to the
empirical one. We establish a relationship between the
single-particle widths obtained and the experimental ones.

The theoretical scheme with a detailed analysis of the various
approaches considered in the evaluation of the spectral function is
presented in Sect.~\ref{sect2}. Here we also show and discuss the
results obtained. We present a systematic study of the scaling
function as well as the momentum distribution considering several
different models and taking into account the role played by FSI. A
summary of the work and our conclusions are presented in
Sect.~\ref{sect3}.

\section{Scaling Function in Relation to the Spectral Function and Momentum Distribution}\label{sect2}

In this section we give the main relationships used in our approach
in order to find a simultaneous description of the spectral
function, momentum distribution and scaling function. As mentioned
in the Introduction, the scaling function is given as a ratio
between the inclusive electron-nucleus inclusive cross section and
the electron-nucleon cross section at $p= |y|$ and ${\cal E}=0$.
Within PWIA the scaling function is expressed in terms of the
spectral function by Eq.~(\ref{scaling_function}). It was shown
in~\cite{23ant10} that in this scheme the equations that relate the
scaling function $F(q,y)$ with the spectral function in the regions
of negative and positive values of the scaling variable $y$ have the
form: \ba {\dfrac{1}{2\pi}}F(q,y) &\!=\!&
\int_{-y}^{Y(q,y)}\!\!p\,dp\int_0^{{\cal E}^-(p;q,y)}d{\cal E}
S(p,{\cal E}) \ \ \ \ \ \ \ \ \ \ \ \ \ \ \ \ \ \ \ \ \ \ \ \ \ \ \
\ \ \ \ \ \ \ \ \ \ \ \ \ \ \mbox{if}\ y<0 \label{eq:Fneg}
\\
{\dfrac{1}{2\pi}}F(q,y) &\!=\!& \int_{0}^y p\,dp\int_{{\cal E}^+(p;q,y)}^{{\cal E}^-(p;q,y)}
d{\cal E} S(p,{\cal E})\!+\!
 \int_{y}^{Y(q,y)}\!\!p\,dp\int_0^{{\cal E}^-(p;q,y)}\!d{\cal E} S(p,{\cal E})
\ \ \ \mbox{if}\ y>0\, . \label{eq:Fpos} \ea In Eqs.~(\ref{eq:Fneg})
and (\ref{eq:Fpos}): \ba y(q,\omega) &=&
\left\{(M_A^0+\omega)\sqrt{\Lambda^2-M_B^{0^2}W^2}-q\Lambda\right\}/W^2,
\label{ysmall}
\\
Y(q,\omega) &=&
\left\{(M_A^0+\omega)\sqrt{\Lambda^2-M_B^{0^2}W^2}+q\Lambda\right\}/W^2,
\label{ylarge}
\ea
\be {\cal
E}^{\pm}(p;q,\omega)=(M_A^0+\omega)-\left[\sqrt{(q\pm p)^2+m_N^2}+
  \sqrt{M_B^{0^2}+p^2}\right],
\ee where $\omega $ is the energy transfer, $M_A^0$ is the target
nuclear mass, $m_N$ is the nucleon mass, $M_B^0$ is the ground-state
mass of the residual nucleus and $\Lambda\equiv
(M_B^{0^2}-m_N^2+W^2)/2$ with $W\equiv\sqrt{(M_A^0+\omega)^2-q^2}$
being the final-state invariant mass.

In the RFG model the dimensionless scaling variable $\psi $ is
introduced~\cite{14scaling88,15ant11,16DS199,17DS299} in the form:
\be \psi =\frac{1}{\sqrt{\xi_F}} \frac{\lambda -
\tau}{\sqrt{(1+\lambda) \tau + \kappa \sqrt{\tau (1+\tau)}}} \, ,
\label{psi-RFG} \ee where $\eta_F =k_F/m_N$, $\xi _F = \sqrt{1+ \eta
_F^2}-1$ is the dimensionless Fermi kinetic energy, $\kappa  =
q/(2m_N)$, $\lambda  = \omega/(2m_N)$, and $\tau  = |Q^2| / (4
m_N^2)= \kappa^2 -{\lambda}^2$ is the dimensionless absolute value
of the squared 4-momentum transfer. The physical meaning of ${\psi
}^2$ is the smallest kinetic energy (in units of the Fermi energy)
that one of the nucleons responding to an external probe can have.

The scaling variables $y$ and $\psi $ are closely
related~\cite{16DS199,17DS299}: \be\label{eq:y_psi} \psi = \left(
\frac{y}{k_F} \right) \left[ 1
+\sqrt{1+\frac{m_N^2}{q^2}}\frac{1}{2} \eta_F \left( \frac{y}{k_F}
\right) +{\cal O} [\eta_F^2] \right] \simeq \dfrac{y}{k_F}\,, \ee
where $\eta_F$ is small, typically $\approx$1/4. The dimensionless
scaling function $f(\psi )$ is introduced ({\it
e.g.}~\cite{17DS299}) in the RFG model: \be\label{eq:f_(psi)}
f_\text{RFG}(\psi) = k_F F_\text{RFG}(\psi)=
\dfrac{3}{4}\left(1-\psi^2\right)\Theta\left(1-\psi^2\right)\,. \ee
As observed, the RFG model leads to a universal scaling function
$f_{RFG}$ which depends only on the scaling variable $\psi$, but
does not depend on the momentum transferred or on the nuclear
species, that is, it superscales.

\subsection[]{Theoretical Spectral Functions: Independent Particle Shell Model and Beyond}

As noted in the Introduction the aim of the present work is to
construct a realistic spectral function that leads to good agreement
with the scaling function obtained from the inclusive
electron-nucleus scattering data. We start with a given form for the
spectral function, {\it viz.} that of the IPSM:
\begin{equation}\label{HF}
    S_{IPSM}(p,{\cal E})=\sum_{i}2(2j_i+1)n_i(p) \delta({\cal E}-{\cal E}_i),
\end{equation}
where $n_i(p)$ is the momentum distribution of the shell-model
single-particle state $i$ and ${\cal E}_i$ is the eigenvalue of the
energy of the state $i$. One may reasonably expect that when effects
beyond mean field are considered, the dependence on the energy here
would be better represented by a function with a finite width in
energy instead of by a $\delta$-function. To explore the possible
effect of a finite energy spread in Eq.~(\ref{HF}), in what follows
the energy dependence in Eq.~(\ref{HF}) (the $\delta $-function) is
usually replaced by a Gaussian distribution $G_{\sigma _i}({\cal E} - {\cal
E}_i)$:
\begin{equation}\label{HF+Gauss}
    S(p,{\cal E})=\sum_{i}2(2j_i+1)n_i(p) G_{\sigma_i}({\cal E}-{\cal E}_i),
\end{equation}
where
\begin{equation}\label{eq3a}
G_{\sigma_i}({\cal E}-{\cal E}_i)=
\dfrac{1}{\sigma_i\sqrt{\pi}}e^{-\frac{({\cal E}-{\cal E}_i)^2}{\sigma_i^2}}
\end{equation}
and $\sigma _i$ is a parameter for a given single-particle state $i$
that is related to the width of the hole state $i$. In the present
work we focus on the case of $^{16}$O, together with a few results
for $^{12}$C. For both nuclei we consider two parameters $\sigma_{1s}$
and $\sigma _{1p}$ that are related to the widths of the $1s$
and $1p$ hole states, respectively. We note that for simplicity we do not
consider the differences in the spin-orbit partners $1p_{3/2 }$ and $1p_{1/2}$ states. We look for a best fit of the
parameters in order to provide a good simultaneous description of
the experimental scaling function $f(\psi)$, the experimental values
for the widths of the hole states and the high-momentum tail of the
momentum distribution.

With a methodical purpose in mind we also consider and use
for a comparison with the Gaussian in Eq. (\ref{eq3a}) another form
of the energy dependence, namely the so-called ``Lorentzian
function'' $L_{\Gamma_i}({\cal E} - {\cal E}_i)$:
\begin{equation}\label{lorent}
L_{\Gamma_i}({\cal E}-{\cal E}_i)=
\dfrac{1}{\pi}\dfrac{\Gamma_i/2}{(E-E_i)^2+(\Gamma_i/2)^2}\, ,
\end{equation}
where $\Gamma_i$ is the width for a given single-particle hole state
$i$.

\begin{figure}[htb]
\begin{center}
\includegraphics[width=86mm]{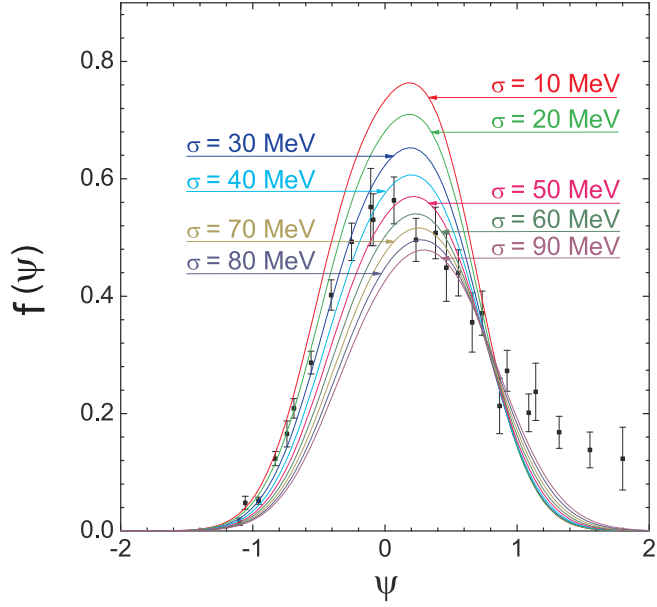}
\caption{(Color online) Results for the scaling function $f(\psi)$
for $^{16}$O obtained using HO single-particle wave functions (for
$\sigma = 10-90$~MeV) are compared with the longitudinal
experimental data. The value of $q$ is fixed to $q=1$ GeV/c.}
\label{fig01}
\end{center}
\end{figure}

\begin{figure}[htb]
\begin{center}
\includegraphics[width=\textwidth]{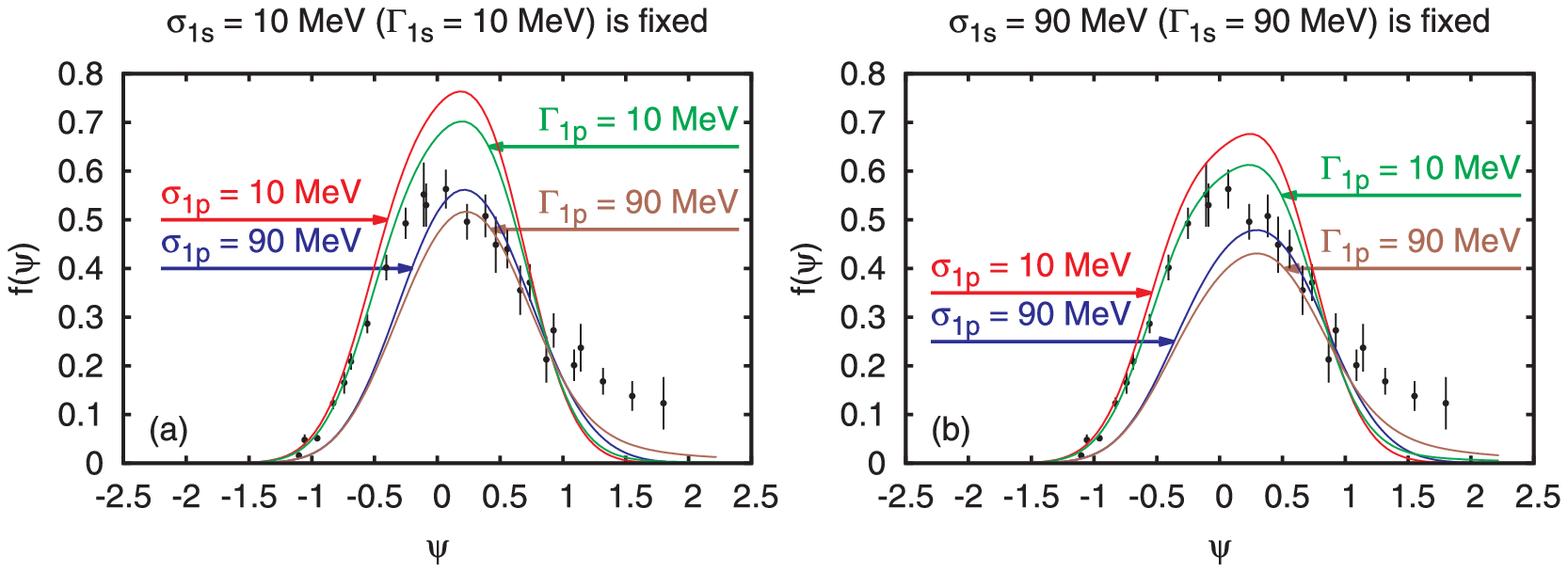}
\caption{(Color online). Results for the scaling function $f(\psi)$
for $^{16}$O  obtained using HO single-particle wave functions. The
values of $\sigma_{1s}$ ($\Gamma_{1s}$ in the case of Lorentzian functions, see Eq.~(\ref{lorent})) are fixed and $\sigma _{1p} = 10$~MeV ($\Gamma_{1p} = 10$~MeV) and $\sigma _{1p} = 90$~MeV ($\Gamma_{1p} = 90$~MeV) have been used. The results are compared with the longitudinal experimental data. The value of $q$ is fixed to $q=1$ GeV/c.} \label{fig02}
\end{center}
\end{figure}

\begin{figure}[htb]
\begin{center}
\includegraphics[width=\textwidth]{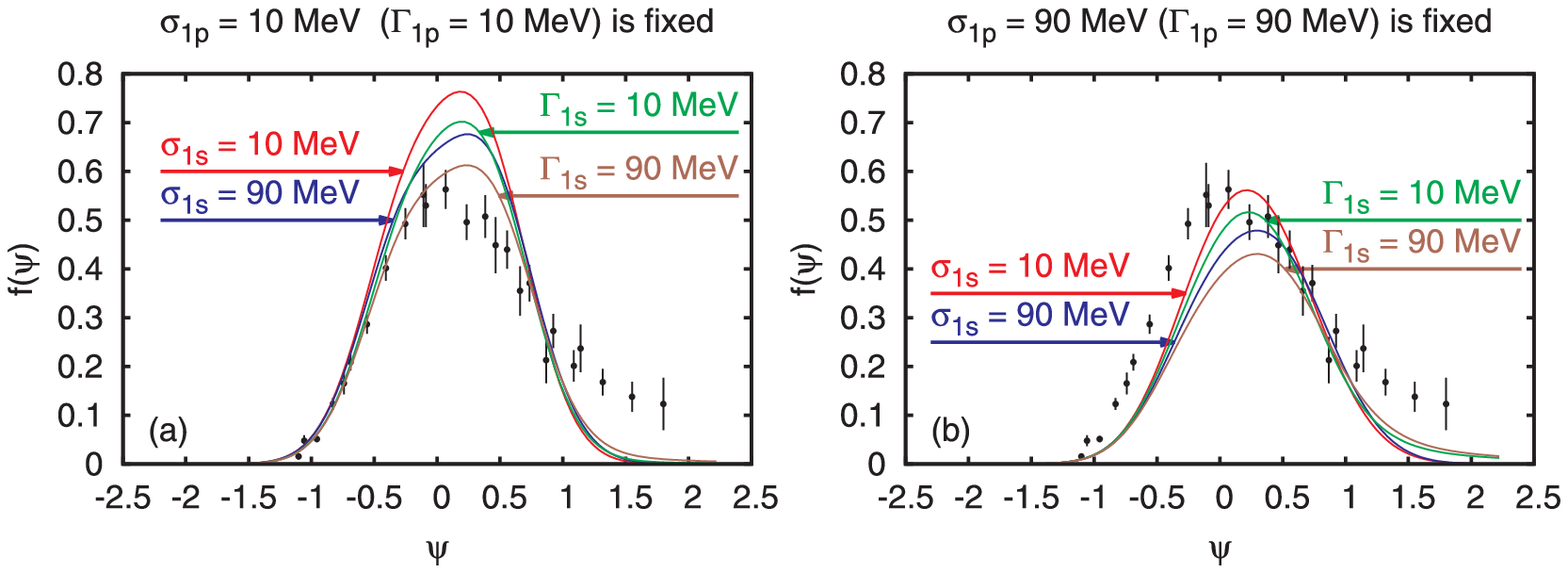}
\caption{(Color online). Results for the scaling function $f(\psi)$
for $^{16}$O  obtained using HO single-particle wave functions.
The values of $\sigma_{1p}$ ($\Gamma_{1p}$) are fixed and $\sigma _{1s} = 10$~MeV ($\Gamma_{1s} = 10$~MeV) and $\sigma _{1s} = 90$~MeV ($\Gamma_{1s} = 90$~MeV) have been used. The results are compared with the longitudinal experimental data. The value of $q$ is fixed to $q=1$ GeV/c.}
\label{fig03}
\end{center}
\end{figure}

We start by taking $n_i(p)$ to be the momentum distribution of the
harmonic-oscillator shell-model single-particle state $i$. Next, as
can be seen from Eqs.~(\ref{HF+Gauss}) and~(\ref{eq3a}) we account
for the effects of nucleon correlations that give widths to the
energy distributions of the hole strengths seen in ($e,e'$) or
($e,e'p$) reactions. These widths may not in fact be symmetric,
although in the present work for simplicity we limit ourselves to
symmetric ones. In Fig.~\ref{fig01} are given the results for the
scaling function compared with the longitudinal experimental data.
There we have assumed equal values of $\sigma_{1s}$ and
$\sigma_{1p}$ ($\sigma_{1s} = \sigma_{1p}\equiv \sigma$) and
vary $\sigma$ in the region $\sigma= 10-90$~MeV. In contrast, results
in Fig.~\ref{fig02} and Fig.~\ref{fig03} correspond to fixed values
of $\sigma_{1s}$ ($\Gamma_{1s}$ in the case of Lorentzian functions, see Eq.~(\ref{lorent})) and $\sigma_{1p}$ ($\Gamma_{1p}$) taken only at the extreme values $10$~MeV and $90$~MeV, and vice versa.

%%%%%%%%%%%%%%%%%%%%%%%%%%%%%%%%%%%%%%%
%%%%%%%%%%%%%%%%%%%%%%%%%%%%%%%%%%%%%%%

%%%%%%%%%%%%%%%%%%%%%%%%%%%%%%%%%%%%%%%
%%%%%%%%%%%%%%%%%%%%%%%%%%%%%%%%%%%%%%%

%%%%%%%%%%%%%%%%%%%%%%%%%%%%%%%%%%%%%%
%%%%%%%%%%%%%%%%%%%%%%%%%%%%%%%%%%%%%%

One can see that at fixed values of the parameter $\sigma_{1s}$  ($\Gamma_{1s}$) and
for running values of $\sigma _{1p}=10$~MeV ($\Gamma_{1p}=10$~MeV) and $90$~MeV the main effects in the scaling function are observed in the shape of the curve and
its maximum value which increases significantly as $\sigma_{1p}$  ($\Gamma_{1p}$)
goes down. Likewise the curve is extended in the negative $\psi$
region. On the other hand, in the case of fixed values of $\sigma
_{1p}$  ($\Gamma_{1p}$) and running values of $\sigma _{1s}= 10$~MeV  ($\Gamma_{1s}=10$~MeV) and $90$~MeV, a smaller decrease of
the maximum is observed and the discrepancies in the extended tail at $\psi<0$ tend to disappear, giving rise to a similar shape for all $\sigma_{1s}$ ($\Gamma_{1s}$) values considered.

The main conclusion from our results presented in
Figs.~\ref{fig01}--\ref{fig03} is that, at least with a symmetric
energy spread for the single-particle energy levels, it is not
possible to get an asymmetry of the longitudinal scaling
function similar to that shown by the data.
%What the energy spread can do is to
%allow a shift of the strength to the region of negative values of
%$\psi$ as illustrated in Figs.~\ref{fig02} and~\ref{fig03}.

Making use of Eq.~(\ref{nk}) we calculate the IPSM momentum
distribution which, as we are using HO single-particle wave
functions, does not present a high-momentum tail. Our next step is
to use natural orbitals (NOs) for the single-particle wave functions
and occupation numbers employing a method where short-range NN
correlations are taken into account. In what follows we use the NO
representation of the one-body density matrix (OBDM) obtained within
the lowest-order approximation of the Jastrow correlation
method~\cite{28Stoitsov93}.

The NOs $\varphi_\alpha (r)$ are defined~\cite{29Lowdin55} as the
complete orthonormal set of single-particle wave functions that
diagonalize the OBDM:
\begin{equation}
\rho (\mathbf{r},\mathbf{r}^{\prime} )=\sum_{a} N_{a} \varphi_{a}^{*}(\mathbf{r}) \varphi_{a}
(\mathbf{r}^{\prime}) ,
\label{defNO}
\end{equation}
where the eigenvalues $N_{\alpha} $ ($0\leqq N_{\alpha}\leqq 1$, $ \sum_{\alpha} N_{\alpha}=A$)
are the natural occupation numbers.

\begin{figure}[htb]
\begin{center}
\includegraphics[width=86mm]{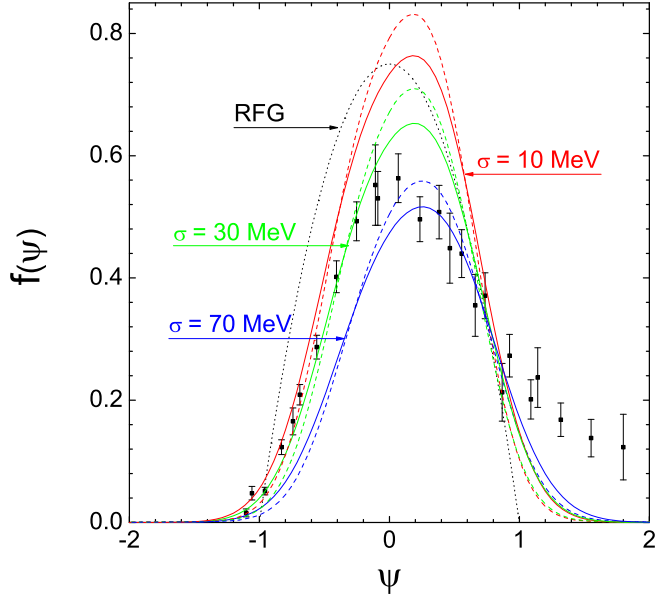}
\caption{(Color online) Results for the scaling function $f(\psi)$
for $^{16}$O obtained using NOs (dashed lines) and harmonic
oscillator (HO) single-particle wave functions (solid lines) for
various values of the parameters $\sigma_{1s}=\sigma_{1p}=10$~MeV
(red lines), 30~MeV (green lines) and 70~MeV (black lines). The RFG
results are shown for comparison (dotted line). The value of $q$ is fixed to
$q=1$ GeV/c.} \label{fig04}
\end{center}
\end{figure}

The NO single-particle wave functions are used to obtain the
momentum distributions $n_i(p)$, and from them the spectral function
according to Eq.~(\ref{HF+Gauss}). The results for the scaling
function obtained using NOs and HO single-particle wave functions
for various values of the parameters $\sigma _{1s}$ and $\sigma_{1p}
$ are given in Fig.~\ref{fig04}. They are represented by solid (HO)
and dashed (NO) lines and compared with the RFG result (dotted)
presented also for reference. As observed, the main effect
introduced by the use of NO, compared with HO, is an enhancement in
the maximum of the order of $\sim$10$\%$. In contrast, the tail is
slightly reduced in the region of negative $\psi$. However, notice
that both models lead to a weak asymmetry in the scaling function
$f(\psi)$ that is not in accordance with the significant tail
extended to positive $\psi$-values, seen in the analysis of $(e,e')$
data. It should be mentioned here that in addition to $1s$- and
$1p$-components there are also $1d$- and $1f$-components in the NOs
obtained from the Jastrow correlation method. Unless specified
otherwise we take the same values of $\sigma $ for all of them.

%%%%%%%%%%%%%%%%%%%%%%%%%%%%%%%%%%%%%%
%%%%%%%%%%%%%%%%%%%%%%%%%%%%%%%%%%%%%%

\begin{figure}[tb]
\begin{center}
\includegraphics[width=86mm]{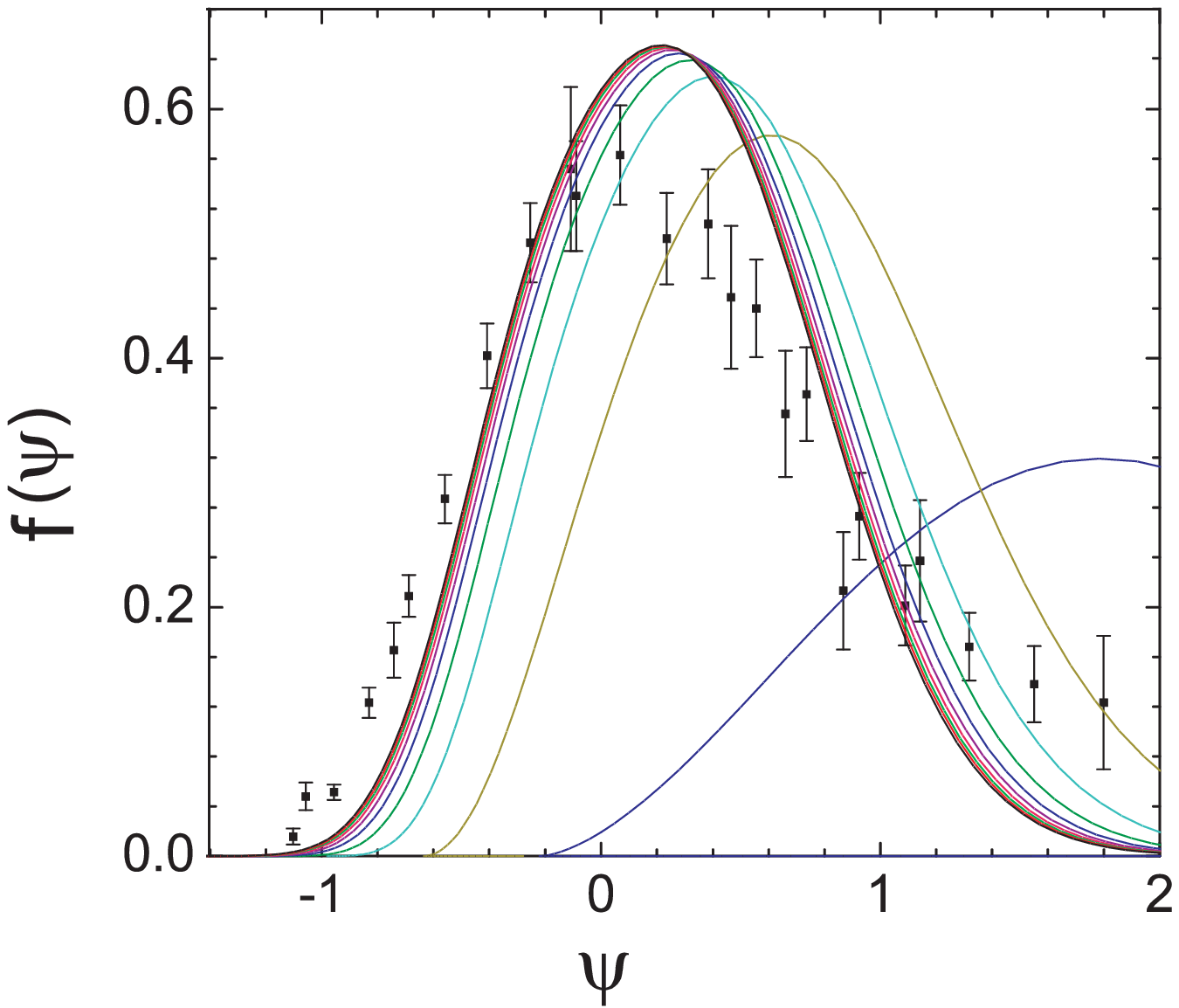}
\caption{(Color online) The scaling function $f(\psi)$ for $^{16}$O
calculated using HO single-particle wave functions (with $\sigma =
0$) is presented as a function of $\psi$ for a range of fixed values
of $q$ extending from 100 MeV/c (right-most curve) to 2000 MeV/c
(left-most curve) in steps of 200~MeV/c. For $q > 700$~MeV/c
scaling of the first kind is seen to be fulfilled.} \label{fig07}
\end{center}
\end{figure}

In Fig.~\ref{fig07} we present the evolution of the scaling function
$f(\psi)$ for different values of $q$ running from $100$ to
$2000$~MeV/c. Results have been obtained making use of the HO
momentum distributions for the $1p$- and $1s$-shells in $^{16}$O. From these one gets the spectral function according to
Eqs.~(\ref{HF+Gauss},~\ref{eq3a}) and finally the scaling function
using the expressions in Eqs.~(\ref{eq:Fneg},~\ref{eq:Fpos}). As
already mentioned, the HO model is not capable of producing the
strong asymmetry observed in the data, but it can be seen that for
$q>600-700$~MeV/c scaling of first kind is fulfilled.

%%%%%%%%%%%%%%%%%%%%%%%%%%%%%%%%%%%%%%
%%%%%%%%%%%%%%%%%%%%%%%%%%%%%%%%%%%%%%

\begin{figure}[htb]
\begin{center}
\includegraphics[width=86mm]{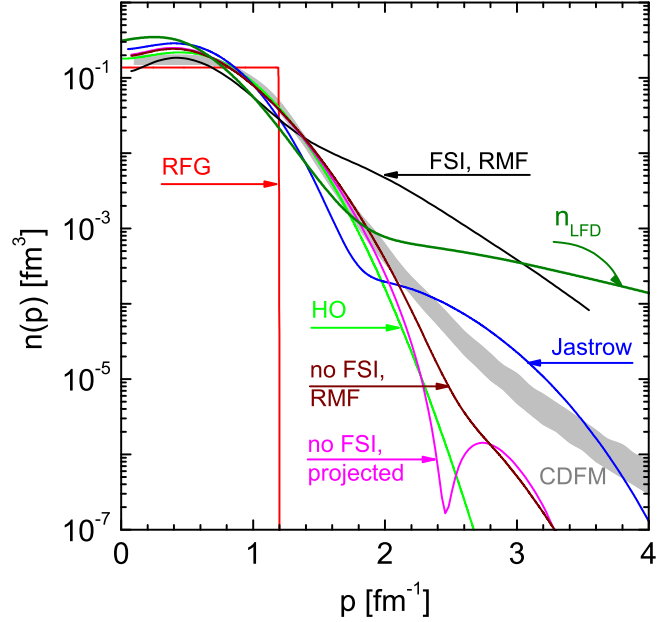}
\caption{(Color online) Results for the momentum distribution for
$^{12}$C obtained using harmonic oscillator single-particle wave
functions (HO -- green line); NO taken from the Jastrow model
(Jastrow -- blue line); Relativistic Fermi Gas (RFG -- red line);
CDFM results (CDFM -- grey area~\cite{20ant05,22ant07}); LFD results
(LFD -- olive line~\cite{22ant07}). Normalization: $\int d\mathbf{k} n(\mathbf{k})=1$. } \label{fig05}
\end{center}
\end{figure}

In what follows we focus on the study of the momentum distribution
obtained in different approaches. Results are presented in
Fig.~\ref{fig05} where we use a log scale in order to emphasize the
differences. The RFG momentum distribution is compared with
the result obtained using HO single-particle wave functions and the
NOs from the Jastrow correlated approach~\cite{28Stoitsov93}. The
nucleon momentum distribution $n^\text{LFD}$ obtained
in~\cite{30ant14,22ant07} by using the Light-Front Dynamics method
(LFD)~\cite{31ant15} is presented in Fig.~\ref{fig05} as well. We
also show in the same figure the results obtained
in~\cite{caballero} within the Relativistic Mean Field (RMF) model
with and without taking into account FSI. Finally, we give in
Fig.~\ref{fig05} as an example (being considered and used in our
previous work) the result obtained with the Coherent Density
Fluctuation Model (CDFM) (\cite{11ant01,04ant02,07ant02a}; see
also~\cite{20ant05,22ant07}). The model is a natural extension of
the RFG model. It is based on the  $\delta$-function limit of the
generator coordinate method~\cite{new_cdfm1} and accounts for
long-range NN correlations of collective type. In it the nucleon
momentum distribution has the form: \be n({\bf k})=\int d{\bf
r}W({\bf r},{\bf k}), \label{cdfm1} \ee where the CDFM Wigner
distribution function is:
\begin{equation}
W({\bf r},{\bf k})=\int_{0}^{\infty}dx|F(x)|^{2} W_{x}({\bf
r},{\bf k})
\label{cdfm2}
\end{equation}
with
\begin{equation}
W_{x}({\bf r},{\bf k})=\frac{4}{(2\pi)^{3}}\Theta (x-|{\bf
r}|)\Theta (k_{F}(x)-|{\bf k}|)
\label{cdmf3}
\end{equation}
and
\begin{equation}
k_{F}(x)=\left[\frac{3\pi^{2}}{2}\rho_{0}(x)\right]^{1/3},\;\;\;\; \rho_{0}(x)=\frac{3A}{4\pi x^{3}}.
\label{cdfm4}
\end{equation}
The model weight function $F(x)$ is obtained by means of a known density distribution $\rho (r)$ for a given nucleus:
\begin{equation}
|F(x)|^{2}=-\frac{1}{\rho_{0}(x)} \left. \frac{d\rho(r)}{dr}\right
|_{r=x}, \;\;\;\;(\mathrm{at} \;\; d\rho(r)/dr\leq 0).
\label{cdfm5}
\end{equation}
The CDFM has been applied to studies of the superscaling phenomenon (e.g.~\cite{19ant04,20ant05,21ant06,22ant07}) in inclusive electron-nucleus scattering, as well as to analyses of neutrino and antineutrino scattering by nuclei of both charge-changing~\cite{22ant07,new_cdfm2} and neutral-current~\cite{new_cdfm3} types.

Before entering into a discussion of the RMF results, it is
interesting to point out the significant differences introduced by
the other models. As noticed, the presence of nucleon correlations
(through the Jastrow approach) leads to a significant tail at high
momentum values compared with the pure HO result. On the contrary,
the momentum distribution is slightly reduced for
small-to-intermediate $p$. As can be seen, the tail at high $p$ is
extremely large for $n^\text{LFD}(p)$ and is also present in the
case of the CDFM model. The latter is a result of the NN correlation
effects accounted for in the model. Later in this work we also show
the role of these correlations within the CDFM
approach~\cite{22ant07} (with phenomenologically introduced FSI
effects) on the scaling function $f(\psi )$ in the cases of $^{12}$C
(Fig.~\ref{fig09}) and $^{16}$O (Figs.~\ref{fig09a}
and~\ref{fig09b}).

%%%%%%%%%%%%%%%%%%%%%%%%%%%%%%%%%%%%%%
%%%%%%%%%%%%%%%%%%%%%%%%%%%%%%%%%%%%%%

Next we discuss in more detail how the momentum distribution
function is evaluated within the general framework of the RMF. The
particular case where FSI are turned on is considered in the next
subsection. In the RMF case without FSI the momentum distribution
$n^{RMF}(p)$ is given in the form: \be
n^{RMF}(p)=\sum_\text{shells}n_\text{shell}(p)\label{RMFwoFSI} \ee
with \be
n_\text{shell}(p)=|\varphi_\text{shell}(p)|^2,\label{nshellbound}
\ee where $\varphi_\text{shell}(p)$ are single-particle 4-component
wave functions that are solutions of the Dirac equation with an RMF
(relativistic Hartree) potential.

It is also interesting to consider the momentum distribution for the case in
which the relativistic bound nucleon wave functions
are projected over positive energy components. %Hence, the projected single-particle wave functions are given by
In this case the four-spinors $\varphi_\text{shell}(p)$ go into
four-spinors $\widetilde{\varphi}_\text{shell}(p)$ by the action of
the positive-energy projection operator \be
\widetilde{\varphi}_\text{shell}(p)= \left(\dfrac{\slash\!\!\!p + m}
{2m}\right)\varphi_\text{shell}(p) \, . \ee Hence the
``positive-energy projected'' momentum distribution results: \be
\widetilde{n}^{RMF}(p)=\sum_\text{shells}\widetilde{n}_\text{shell}(p)
=\sum_\text{shells}|\widetilde{\varphi}_\text{shell}(p)|^2 \, .
\label{nonrelRMF} \ee

The latter from Eq.~(\ref{nonrelRMF}) is labeled in Fig.~\ref{fig05}
as {\sl ``no FSI, projected''}, although we should point out that it
differs from the usual non-relativistic analyses where two-component
Schr\"odinger-like equations are considered and a non-relativistic
reduction of the relativistic $4\times 4$ current operator into a
$2\times 2$ form is used. Here, we project out the negative-energy
components of the bound nucleon wave functions while a fully
four-component description of the Dirac spinors is maintained. The
outgoing nucleon is described by means of a relativistic (Dirac)
plane wave. These two assumptions, {{\it i.e.,} plane waves in the
final state and positive-energy projection in the initial one, lead
to the so-called Plane-Wave Impulse Approximation (PWIA), which
should be distinguished from the Relativistic Plane-Wave Impulse
Approximation (RPWIA) where the plane-wave approach for the ejected
nucleon is also considered, but where the fully relativistic bound
nucleon wave function (without projection) is used.

In PWIA the exclusive electron ($e,e'N$) cross section factorizes in the form:
\be
\left.\dfrac{d\sigma}{d\Omega'd\epsilon'd\Omega_N
%d\epsilon_N
}\right|_\text{PWIA}=
K\sigma^{eN}\widetilde{n}_\text{shell}(p)\, , \label{exccs}
\ee
where $K$ is a kinematical factor and $\sigma^{eN}$ is the electron-nucleon cross section.
Thus Eq.~(\ref{exccs}) makes it possible to obtain the momentum distribution for different shells
by means of the cross section calculated within the PWIA:
\be
\widetilde{n}_\text{shell}(p)=\dfrac{\left[\dfrac{d\sigma}{d\Omega'd\epsilon'd\Omega_N%d\epsilon_N
} \right]_\text{PWIA}}{K\sigma^{eN}}\, . \label{npcsexpwia} \ee This
result coincides with the one given through the positive-energy
projected wave functions in Eq.~(\ref{nonrelRMF}).

Results for $n^{RMF}(p)$ (labeled as ``no FSI, RMF'') and
$\widetilde{n}^\text{RMF}(p)$ (no FSI, projected) are compared in
Fig.~\ref{fig05}. As shown, $\widetilde{n}^\text{RMF}(p)$ follows
the HO result closely, showing a steep slope with increasing $p$.
Likewise, $n^\text{RMF}(p)$ presents a similar behavior but without
the diffraction minimum due to the presence of the lower components
in the relativistic bound nucleon wave functions. These results confirm
previous studies presented in~\cite{JAC}.

\subsection{{The role of FSI}}

The factorized form of the $(e,e'N)$ cross section shown in
Eq.~(\ref{exccs}) does not apply when final-state interactions are
included in the description of the reaction mechanism. However, one
can introduce a function given as the ratio between the differential
cross section and the single-nucleon cross section (multiplied by
the kinematical factor $K$). This is usually called the reduced
cross section or distorted momentum distribution: \be
\rho(p)=n^\text{dist}(p)=\dfrac{\left[\dfrac{d\sigma}{d\Omega'd\epsilon'd\Omega_N%d\epsilon_N
}
    \right]_\text{FSI}}{K\sigma^{eN}} \, .
\label{npcsexfsi}
\ee

Although in Eq.~(\ref{npcsexfsi}) we show the function
$n^\text{dist}$ to be only dependent on the missing momentum $p$,
this is not so in general because of the non-factorized form of the
$(e,e'N)$ differential cross section when FSI are included. The
distorted momentum distribution depends not only on the missing
momentum $p$ but also on the other independent variables in the
scattering process, for example the momentum transfer $q$. Hence,
calculations for different kinematical situations may lead to
slightly different distorted momentum distributions. However, this
dependence has been proven to be weak in most
cases~\cite{Udias1,Udias2,Udias3}. The resulting effective momentum
distribution $n^\text{dist}(p)$
for a momentum transfer of 1~GeV/c and the momentum of
the final nucleon equal to the momentum transfer
is also displayed in Fig.~\ref{fig05} (labeled as ``FSI, RMF''). In
computing this effective momentum distribution using
Eq.~(\ref{npcsexfsi}), the same real scalar and vector potentials
that describe the bound nucleons are employed to describe the final
nucleon. The nonlinear effective interaction NLSH~\cite{Udias4} has
been used.

As observed, FSI evaluated within the RMF approach gives rise to the presence of a very significant
tail at high missing momentum which is several orders of magnitude
larger than the results obtained within the plane-wave approach.
On the contrary, $n^\text{dist}(p)$ is smaller for $p$-values below the
Fermi momentum. It is important to point out that the right
amount of asymmetry in the longitudinal scaling function required by $(e,e')$ data
is also reproduced by a semi-relativistic shell model when the continuum states
are described with the Dirac-equation-based (DEB) equivalent potential to the RMF
employed here~\cite{Amaro07,Amaro10}.

Once the distorted RMF momentum distribution has been obtained
[Eq.~(\ref{npcsexfsi})], we calculate the ``distorted'' spectral
function $S^\text{dist}(p,{\cal E})$ using Eqs.~(\ref{HF+Gauss}) and
(\ref{eq3a}). Finally, the corresponding scaling and superscaling
functions can be evaluated making use of the procedure outlined
through Eqs.~(\ref{sscalingf},~\ref{eq:Fneg},~\ref{eq:Fpos}). The
results for $f(\psi)$ are given for different values of $\sigma =
\sigma _{1s}=\sigma _{1p}$ (from $10$ to $100$~MeV) in
Fig.~\ref{fig06}. Also, for completeness, we show in
Fig.~\ref{fig06} the scaling function $f(\psi)$ obtained within the
RMF+FSI model, but evaluated directly from the inclusive $(e,e')$
differential cross section~\cite{caballero}, that is, \be
f^\text{dist}(\psi)=k_F
\dfrac{\left[\dfrac{d\sigma}{d\epsilon'd\Omega'}\right]_\text{(e,e')}^\text{RMF+FSI}}
 {\overline{\sigma}^{eN}(q,\omega;p=|y|,{\cal E}=0)}\,. \label{scaling1}
\ee

Results in Fig.~\ref{fig06} show that the RMF with FSI included
gives rise to an asymmetrical scaling function which follows the
behavior of data, irrespective of whether we use
Eqs.~(\ref{HF+Gauss},\ref{npcsexfsi}) or we use
Eq.~(\ref{scaling1}). In the first case, significant discrepancies
emerge when different values of the $\sigma$-parameter are used in
the Gaussian energy-dependent functions. The scaling function
obtained from Eq.~(\ref{scaling1}) is close to that obtained from
Eqs.~(\ref{HF+Gauss},\ref{npcsexfsi}) for $\sigma=30$~MeV. We note
that for the negative $\psi$ region, the scaling function obtained
from the RMF calculation for the cross section lies well below the
data. The comparison with data in this region is thus improved when
a finite width for the energy is introduced in the spectral
function. As observed, the case $\sigma=30$ MeV is the one that
agrees better with experiment and with the theoretical curve
obtained directly from the inclusive cross section (see
Eq.~(\ref{scaling1})). We use
$\sigma=\sigma_{1s}=\sigma_{1p}$ to check the dependence of the
scaling function $f(\psi)$ on $\sigma$. Below we
discuss the optimal values of these parameters
$\sigma_{1s}\ne\sigma_{1p}$, and check the results for the
scaling functions when these parameters give widths close to the
experimental ones.

%%%%%%%%%%%%%%%%%%%%%%%%%%%%%%%%%%%%%%
\begin{figure}[hb]
\begin{center}
\includegraphics[width=86mm]{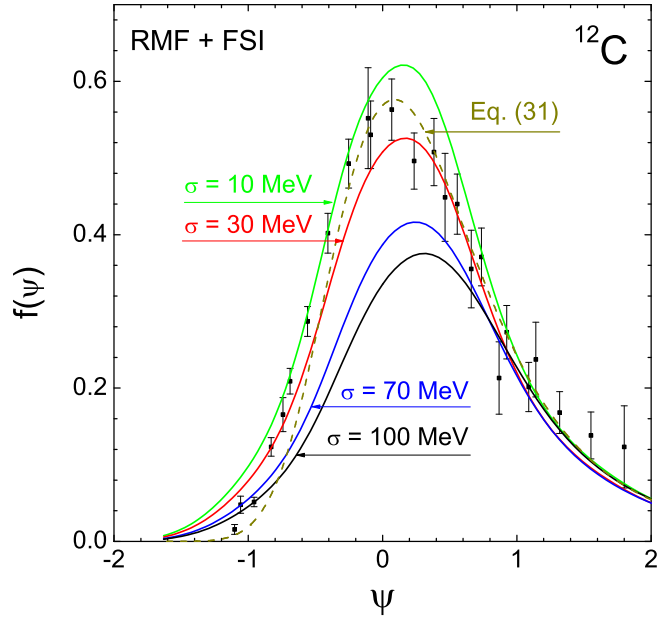}
\caption{(Color online) Results for the scaling function $f(\psi)$
obtained using single-particle wave functions taken from RMF + FSI
for calculation of $n_i(p)$ and for various values of the parameters
$\sigma=\sigma_{1s}=\sigma_{1p}$. Also shown
are results obtained through Eq.~(\ref{scaling1}). Momentum transfer fixed to
$q=1$ GeV/c.} \label{fig06}
\end{center}
\end{figure}
%%%%%%%%%%%%%%%%%%%%%%%%%%%%%%%%%%%%%%

%%%%%%%%%%%%%%%%%%%%%%%%%%%%%%%%%%%%%%
\begin{figure}[htb]
\begin{center}
\includegraphics[width=86mm]{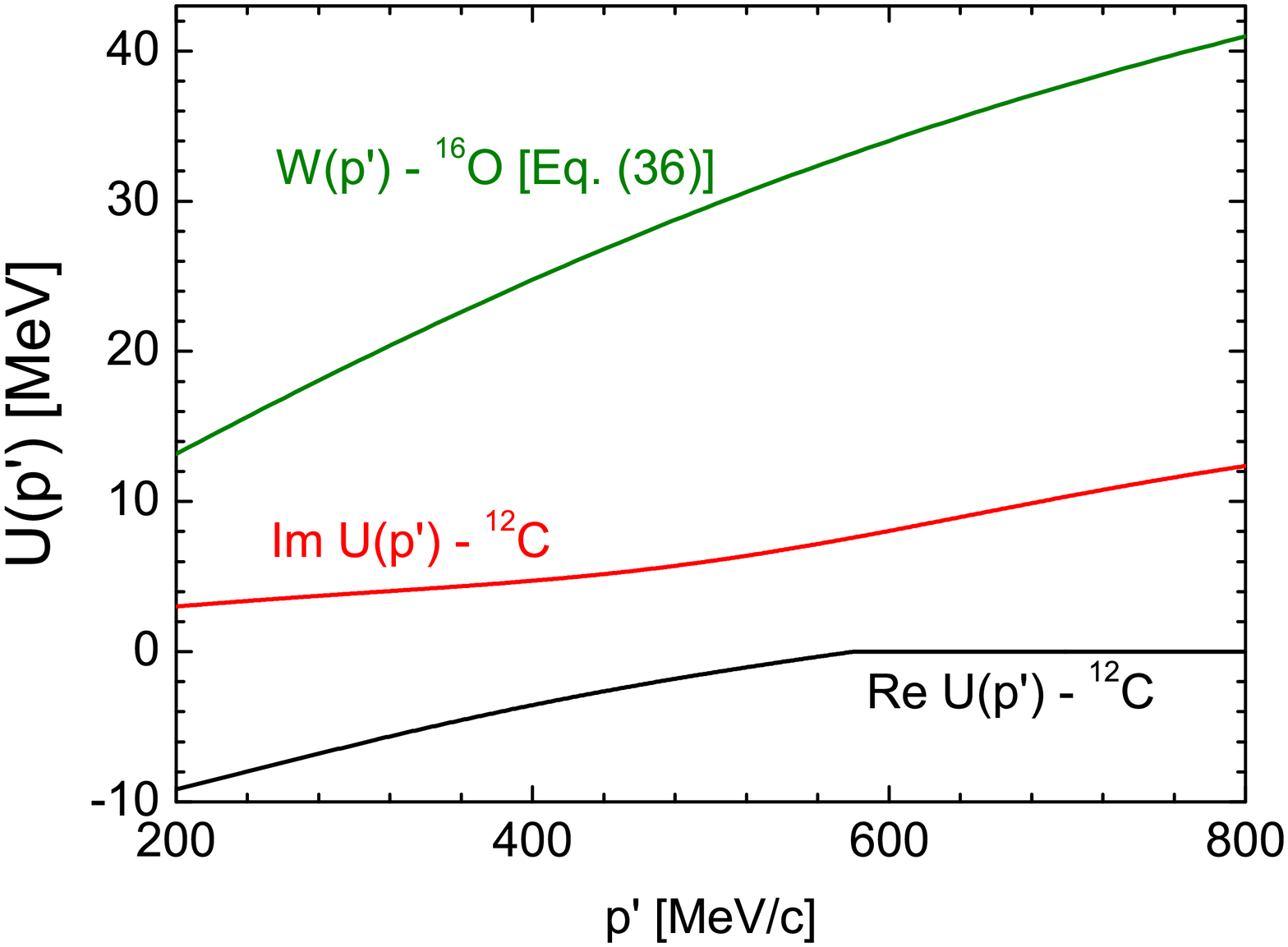}
\caption{(Color online) The real and imaginary parts of the optical
potential $U(p')$ for $^{12}$C calculated by Eq.~(\ref{eq7}) using a
Dirac OP from~\cite{37Cooper1993} and the imaginary part of the optical potential
for $^{16}$O given by Eq.~(\ref{eq9}).} \label{fig08}
\end{center}
\end{figure}
%%%%%%%%%%%%%%%%%%%%%%%%%%%%%%%%%%%%%%

In addition to the description of FSI within the framework of the
RMF model, we have also considered alternative analyses of
final-state interactions in the case of inclusive electron-nuclei
processes. In what follows we discuss a method proposed in
\cite{33ankowski} and present the results we obtained with this
method.

According to the authors of \cite{33ankowski} two types of FSI
effects, Pauli blocking and the interaction of the struck nucleon
with the spectator system described by means of the time-independent
optical potential (OP)
%Following ~\cite{33ankowski}, the interaction of
%the struck nucleon with the spectator system is described by means of the
%time-independent optical potential (OP):
  \begin{equation}\label{eq3}
    U=V-\imath W
    \end{equation}
proposed in \cite{34Horikawa}, can be accounted for by replacing in
the PWIA expression for the
inclusive electron-nucleus cross section %(see~\cite{35Nakamura})}
    \begin{equation}\label{eq5}
        \frac{d\sigma_t}{d\omega d\n q}={2\pi\alpha^2}\frac{\n q}{E_{\ve k}^2}
        \int dE\:d^3p\:\frac{S_t(\ve p, E)}{E_{\ve p}E_{\ve {p'}}}%\times
        \delta\big(\omega+M-E-E_\ve{p'}\big)L_{\mu\nu}^\text{em}H^{\mu\nu}_{\text{em, }t}\, .
    \end{equation}
the energy-conserving delta-function by
    \begin{equation}\label{eq4}
        \delta (\omega+M-E-E_\ve{p'}) \rightarrow \dfrac{W/\pi}{W^2+[\omega+M-E-E_\ve{p'}-V]^2}.
    \end{equation}
In Eq.~(\ref{eq5}) the index $t$ denotes the nucleon isospin, $L_{\mu\nu}^\text{em}$ is the
leptonic tensor, $H^{\mu\nu}_{\text{em, }t}$ is the hadronic tensor and $S_t(p,E)$ is the proton
(neutron) spectral function. The terms $E_{\ve k}$, $E_{\ve p}$, $E_{\ve {p'}}$ and $E$
represent the initial electron energy, the energy of the nucleon inside the nucleus,
the ejected nucleon energy and the removal energy, respectively (see~\cite{33ankowski} for
details).

The real and imaginary parts of the OP in Eqs.~(\ref{eq3},\ref{eq4})
are obtained from the Dirac OP in \cite{37Cooper1993}. We use
spatially averaged values, evaluating them at the $r$-values that
match their respective root mean squared radii \cite{37Cooper1993}.
As a result the OP $U(p')$ related to the scalar ($S$) and vector
($V$) part of the potential in \cite{37Cooper1993} is obtained in
the form (see also \cite{33ankowski}):
    \begin{equation}\label{eq7}
    E_\ve{p'}+U(\ve{p'})=\sqrt{[M+S(T_\ve{p'},\bar r_S)]^2+\ve{p'}^2}+V(T_\ve{p'},\bar r_V).
    \end{equation}

%%%%%%%%%%%%%%%%%%%%%%%%%%%%%%%%%%%%%%
\begin{figure}[htb]
\begin{center}
\includegraphics[width=86mm]{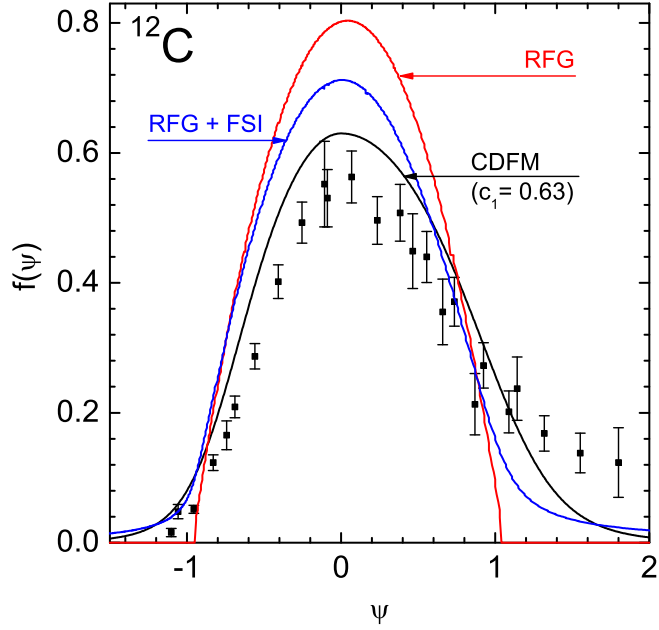}
\caption{(Color online) Results for the scaling function $f(\psi)$
with and without taking into account FSI in the RFG model (for a
given momentum transfer $q=1$~GeV/c and energy of the initial
electron $\epsilon=1$~GeV) are compared with the longitudinal
experimental data and CDFM results using $c_1=0.63$ \cite{22ant07}.
} \label{fig09}
\end{center}
\end{figure}
%%%%%%%%%%%%%%%%%%%%%%%%%%%%%%%%%%%%%%

%%%%%%%%%%%%%%%%%%%%%%%%%%%%%%%%%%%%%%
\begin{figure}[htb]
\begin{center}
\includegraphics[width=86mm]{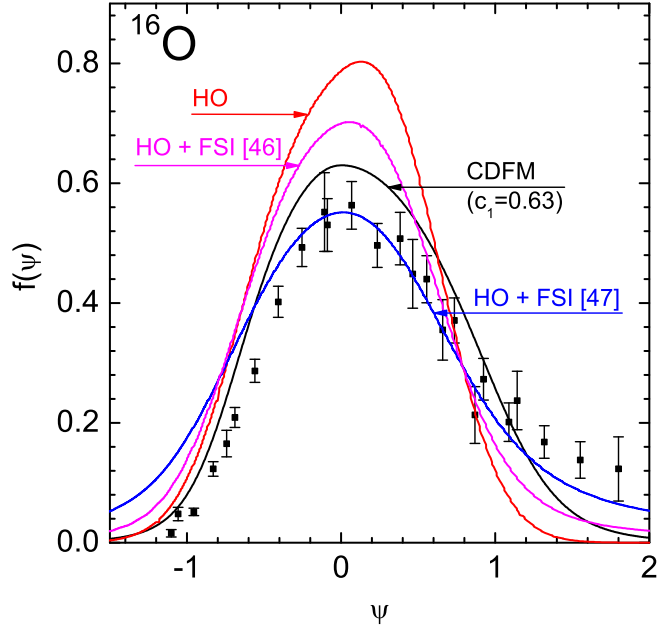}
\caption{(Color online) Results for the scaling function $f(\psi)$
with and without taking into account FSI using the spectral function
in the IPSM~(Eq.~(\ref{HF})) with HO single-particle wave functions
(for a given momentum transfer $q=1$~GeV/c and energy of the initial
electron $\epsilon=1$~GeV) are compared with the longitudinal
experimental data and CDFM results using $c_1=0.63$ \cite{22ant07}.}
\label{fig09a}
\end{center}
\end{figure}
%%%%%%%%%%%%%%%%%%%%%%%%%%%%%%%%%%%%%%

%%%%%%%%%%%%%%%%%%%%%%%%%%%%%%%%%%%%%%
\begin{figure}[htb]
\begin{center}
\includegraphics[width=86mm]{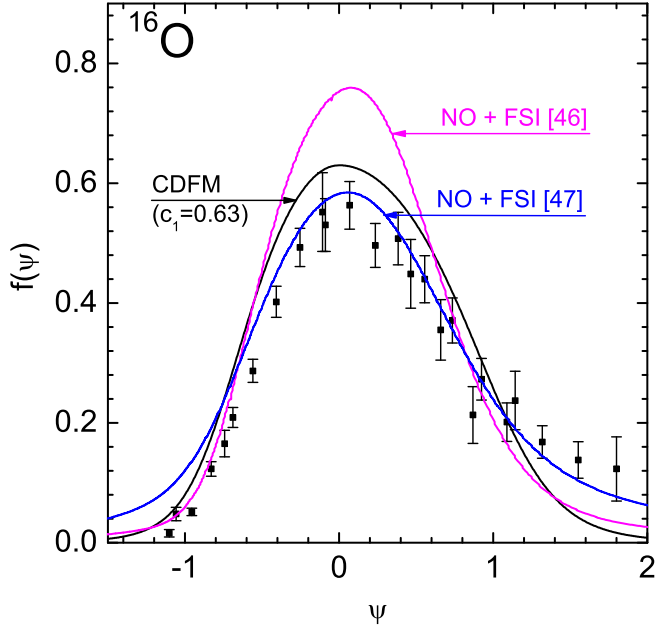}
\caption{(Color online) Results for the scaling function $f(\psi)$
taking into account FSI using NOs from the Jastrow correlation
method (for a given momentum transfer $q=1$~GeV/c, energy of the
initial electron $\epsilon=1$~GeV, and parameters
$\sigma_{1s}=8.7$~MeV,
$\sigma_{1p}=\sigma_{1d}=\sigma_{1f}=0.5$~MeV) are compared with the
longitudinal experimental data and CDFM results using $c_1=0.63$
\cite{22ant07}.} \label{fig09b}
\end{center}
\end{figure}
%%%%%%%%%%%%%%%%%%%%%%%%%%%%%%%%%%%%%%

%%%%%%%%%%%%%%%%%%%%%%%%%%%%%%%%%%%%%%
\begin{figure}[htb]
\begin{center}
\includegraphics[width=86mm]{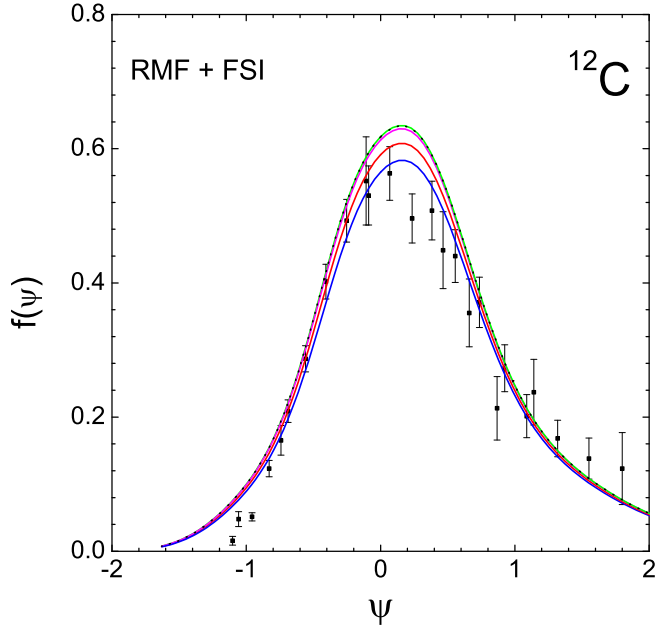}
\caption{(Color online) Results for the scaling function $f(\psi)$
obtained using single-particle wave functions taken from RMF + FSI
for calculation of $n_i(p)$. Five curves are shown. Four (given by solid lines) correspond to
$\sigma_{1s}=7.8$~MeV taken from~\cite{38Jacob1973} and $\sigma_{1p}$
ranging from $5$ MeV (green line) to 20 MeV (blue line) in steps of 5 MeV. The fifth curve (black dotted line)
corresponds to widths for $1s$- and
$1p$-states taken from \cite{dutta} (FWHM$_{1s}=20$~MeV and
FWHM$_{1p}=6$~MeV). Momentum transfer: $q=1$ GeV/c.} \label{fig10}
\end{center}
\end{figure}
%%%%%%%%%%%%%%%%%%%%%%%%%%%%%%%%%%%%%%

We consider also an OP with the imaginary part of the potential $U(p')$ given in~\cite{35Nakamura}:
\begin{equation}\label{eq9}
W=\dfrac{\hbar c}{2}\rho_\text{nucl}\sigma_{NN}\dfrac{|\mathbf{p}'|}{E_{\mathbf{p}'}} \, ,
\end{equation}
where the values of $\rho_\text{nucl}$ and $\sigma_{NN}$ for $^{16}$O are taken to be:
$\rho_\text{nucl}= 0.16$~fm$^{-3}$ and $\sigma_{NN}= 40$~mb.

In Fig.~\ref{fig08} we give the calculated real and imaginary parts
of the OP $U(p')$ from Eq.~(\ref{eq7}) and the imaginary part of the
OP from Eq.~(\ref{eq9}). As noted in~\cite{33ankowski}, for
$|\mathbf{p}'| > 580$~MeV/c the real part of $U(\mathbf{p}')$ is
positive, which is inconsistent with correlated Glauber
theory~\cite{46Benhar}. Therefore, following~\cite{33ankowski} when
$|\mathbf{p}'| > 580$~ MeV/c the real part of the OP is set to zero.
As was noted in Ref.~\cite{46Benhar} the real and the imaginary part
of the optical potential quantitatively have very different effects,
and describe different aspects of the FSI. The imaginary part $W$
accounts for two-body scattering processes involving large momentum
transfers, which lead to a strong damping of the motion of the
recoiling particle, whereas the real part $U$ produces a shift of
its energy. The effect of the imaginary part is known to be dominant
at large momentum.

Following the approach of \cite{33ankowski} we calculate the scaling
function according to Eq.~(\ref{scaling}), but evaluating the cross
section in the numerator by accounting for FSI using the Dirac OP,
{\it i.e.,} exchanging the $\delta $-function in Eq.~(\ref{eq5}) as
shown in Eq.~(\ref{eq4}). The expression for $S(p,{\cal E})$ in
Eq.~(\ref{eq5}) is taken in the form in Eq.~(\ref{HF+Gauss}), where
we use momentum distributions $n_i(p)$ obtained in different
approaches. The results in the case of the RFG model spectral
function are presented in Fig.~\ref{fig09} for a given momentum
transfer $q=1$~GeV/c and energy of the initial electron
$\epsilon=1$~GeV with and without accounting for FSI. The scaling
function $f(\psi)$ using the $S_{IPSM}(p,{\cal E})$ of
Eq.~(\ref{HF}) with HO single-particle wave functions is presented
in Fig.~\ref{fig09a} for $q=1$~GeV/c and $\epsilon=1$~GeV. In this
case we consider two different types of the time-independent optical
potential for $^{16}$O: i) the one obtained by
Eqs.~(\ref{eq3},\ref{eq4},\ref{eq7}) using the scalar and vector
part of the potential from \cite{37Cooper1993} and ii) using the
imaginary part of the potential $U(\mathbf{p}')$ from Eq.~(\ref{eq9}) \cite{35Nakamura}.
%given in
%\cite{35Nakamura}
%    \begin{equation}\label{eq9}
%    W=\dfrac{\hbar c}{2}\rho_\text{nucl}\sigma_{NN}\dfrac{|\mathbf{p}'|}{E_{\mathbf{p}'}} \, ,
%    \end{equation}
%where the values of $\rho_\text{nucl}$ and $\sigma_{NN}$ for $^{16}$O are taken to be:
%$\rho_\text{nucl}= 0.16$~fm$^{-3}$ and $\sigma_{NN}= 40$~mb.

As can be seen from Figs.~\ref{fig09} and~\ref{fig09a} FSI lead to a
decrease in the maximum of the scaling function, while an
enhancement of the tails is observed for both negative and positive
values of $\psi$. Also from Fig.~\ref{fig09a} can be seen the
important role of the type of the time-independent optical potential
used.

In Figs.~\ref{fig09} and~\ref{fig09a}, as well as in
Fig.~\ref{fig09b} we give for a comparison the results for the
scaling function obtained within the CDFM approach~\cite{22ant07} in
which FSI effects are introduced phenomenologically. As can be seen,
accounting for FSI together with the effects of NN correlations is
necessary for a reasonable explanation of the experimental data for
the longitudinal scaling function, especially the maximum and the
tail of $f(\psi )$ at $\psi  > 1$.

Finally we consider results that correspond to a spectral function
constructed with NOs from the Jastrow correlation method, within
this alternative way to take into account FSI. These results are
presented in Fig.~\ref{fig09b} for $q=1$~GeV/c and electron beam
energy $\epsilon=1$~GeV (we consider two different time-independent
optical potentials for $^{16}$O, the same as in Fig.~\ref{fig09a}).
The value of the parameter $\sigma _{1s}$ is fixed to be $8.7$~MeV
(that corresponds to the experimental width of the $1s$-state in
$^{16}$O taken from~\cite{38Jacob1973}) and
$\sigma_{1p}=\sigma_{1d}=\sigma_{1f}=0.5$~MeV. The results depend
very weakly on the choice of parameters $\sigma_i$.

The next step of our work was to obtain optimal values of the
parameters $\sigma_i$ in the Gaussians $G_{\sigma_i}({\cal E}-{\cal
E}_i)$ [Eq.~(\ref{eq3a})] by an additional fitting procedure. These
values of $\sigma _i$ are compared with the experimental widths of
the hole states. The results of our calculations in the case of the
RMF model accounting for the FSI's are given in Fig.~\ref{fig10}.
Here the value of the parameter $\sigma _{1s}$ is fixed to be
$7.8$~MeV (that corresponds to the experimental width of the
$1s$-state in $^{12}$C taken from~\cite{38Jacob1973}) and the values of $\sigma
_{1p}$ are varied. In Fig.~\ref{fig10} is given also the result for
the scaling function obtained when more recent experimental data for
the Full Width at Half Maximum (FWHM) for $1s$- and $1p$-states
(taken from \cite{dutta}) are used. Notice that the latter result (black dotted line) almost coincides with the case $\sigma_{1p}=5$~MeV and $\sigma _{1s}=7.8$~MeV. As can be seen, the results for
the scaling function cannot reproduce very well the experimental
data using experimental widths.

%%%%%%%%%%%%%%%%%%%%%%%%%%%%%%%%%%%%%%
\begin{figure}[htb]
\begin{center}
\includegraphics[width=86mm]{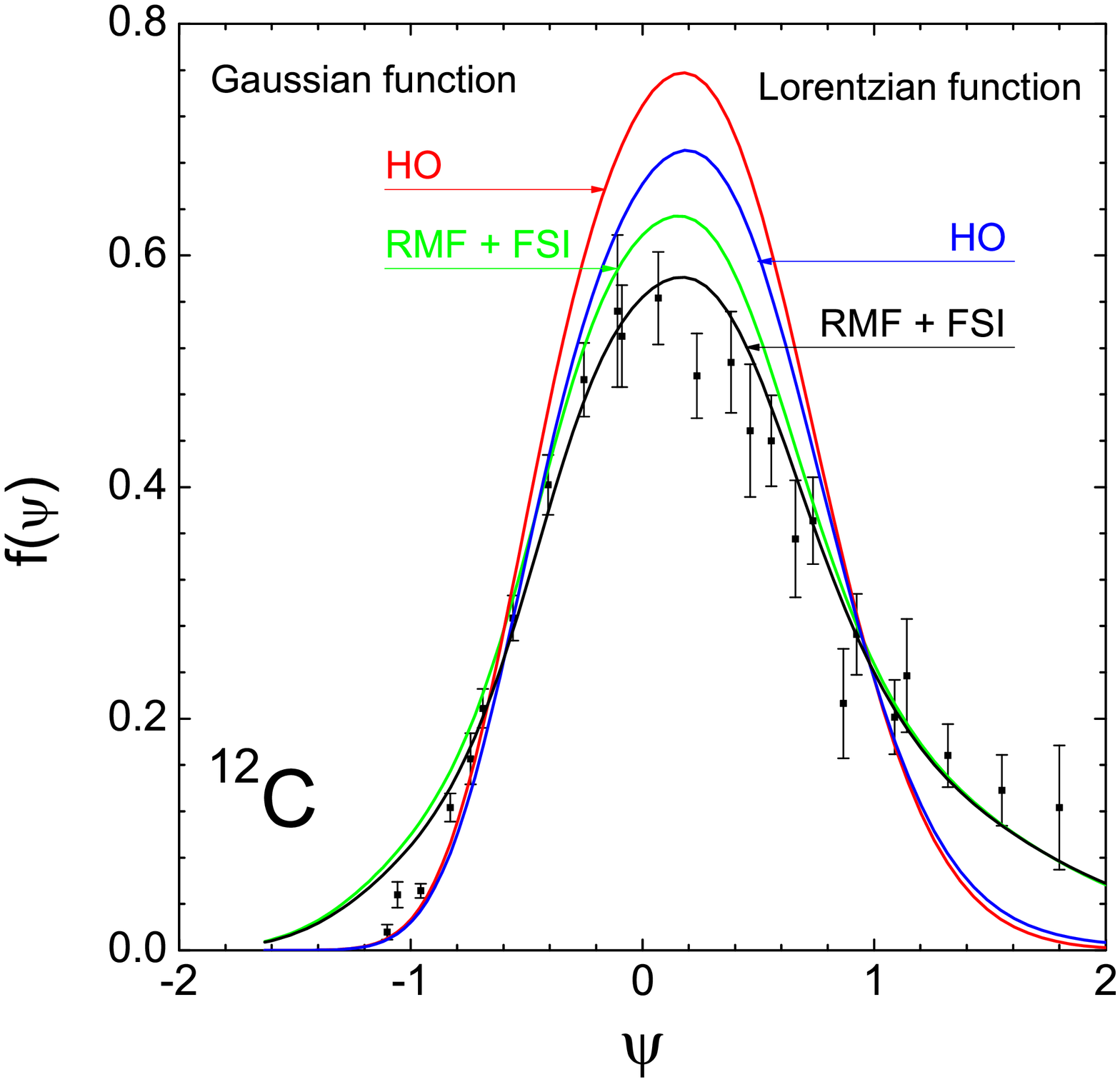}
\caption{(Color online) Results for the scaling function $f(\psi)$
obtained using single-particle wave functions taken from RMF + FSI
for calculation of $n_i(p)$ (parameters $\Gamma _{1p} = 6$~MeV,
$\Gamma _{1s} = 20$~MeV are fixed to the experimental widths of the
$1p$- and $1s$-states in $^{12}$C \cite{dutta}) and two
energy-dependence functions, Lorentzian and Gaussian.} \label{fig11}
\end{center}
\end{figure}
%%%%%%%%%%%%%%%%%%%%%%%%%%%%%%%%%%%%%%

Finally,
%the energy dependence shown in Eq.~(\ref{HF}) (the $\delta $-function) is replaced by the so-called ``Lorentzian function''
%$L_{\Gamma_i}({\cal E} - {\cal E}_i)$:
%\begin{equation}\label{lorent}
%L_{\Gamma_i}({\cal E}-{\cal E}_i)=
%\dfrac{1}{\pi}\dfrac{\Gamma_i/2}{(E-E_i)^2+(\Gamma_i/2)^2}\, ,
%\end{equation}
%where $\Gamma_i$ is the width for a given single-particle hole state
%$i$. In
in Fig.~\ref{fig11} are presented and compared the results for the scaling
function $f(\psi)$ calculated using single-particle functions taken
from HO and RMF+FSI and for two energy-dependence functions,
Lorentzian and Gaussian. We note the good agreement of $f(\psi)$
(using widths close to the experimental $\Gamma _{1p} = 6$~MeV,
$\Gamma _{1s} = 20$~MeV and single-particle functions taken from
RMF+FSI), including the maximum value and the asymmetry of the
scaling function. In this way, together with the results for the
momentum distribution and the scaling function, we obtain a
completion and self-consistency of our study.

\section[]{Conclusions\label{sect3}}

The analysis of the world $(e,e')$ data has clearly demonstrated the
validity of scaling arguments. Not only has the scaling phenomenon
been proven to be fulfilled, but also the specific shape of the
scaling function, with a significant tail extending to high values
of the scaling variable, has emerged from the separated $(e,e')$
longitudinal data. These results constitute a strong constraint for
any theoretical model describing electron scattering reactions.
Moreover, the scaling function is an observable which has been used
in the past in order to get information on the spectral function
$S(p,{\cal E})$ and/or the nucleon momentum distributions of
nucleons in nuclei $n(p)$. However, as explored in detail in our
previous work \cite{23ant10}, the connection between the scaling
function (given directly from the analysis of $(e,e')$ data) and
$S(p,{\cal E})$ or $n(p)$ only exists under very restrictive
conditions: i) the Plane-Wave Impulse Approximation (PWIA) in the
description of the reaction mechanism, and ii) additional
assumptions on the integration limits consistent with the
kinematically allowed region. Being aware of these restrictions and
with some caution in the discussion of results, the link of the
spectral function/momentum distribution with scattering observables
remains a delicate issue in nuclear physics. Effects beyond the
PWIA, such as final-state interactions (FSI), meson-exchange
currents, rescattering processes, {\it etc.,} which may have very
important roles to play in describing $(e,e')$ processes, should be
incorporated in the general analysis.

In this paper we extend the previous work presented in
\cite{23ant10} by considering several models dealing with the
nuclear structure, and including as well different descriptions to
account for FSI. From the momentum distributions provided by three
nuclear models: Harmonic Oscillator, Natural Orbitals with
Jastrow-type correlations and Relativistic Mean Field, we construct
the spectral functions where the energy-dependence given through the
delta function has been changed by introducing Gaussian
distributions. In this way, we present a systematic analysis for
different values of the Gaussian distribution parameters related to
the width of the hole states. Finally, the scaling function is
evaluated by making use of the general derivative expressions that
connect it with the spectral function. We have explored in detail
the role played by Gaussian or Lorentzian energy dependences in the scaling
function showing that, although significant differences may emerge,
all models based on PWIA lead to scaling functions which do not show
the asymmetry that data exhibit, with an extended tail at high
positive values of the scaling variable $\psi$. At the same time we
note that the width in energy introduced for the hole state, as
opposed to the $\delta$-function of the strict IPSM, can however
improve the agreement with data in the negative $\psi$ region of the
scaling function.

Effects of FSI have been proven to be essential in the description
of electron scattering reactions. Hence, we have included FSI in our
analysis within the framework of the RMF and the method using a
complex optical potential \cite{33ankowski}. In the former we obtain
the reduced cross section (or effective distorted momentum
distribution) for each shell as the ratio between the exclusive
$(e,e'N)$ cross section evaluated within the RMF model, and FSI
included, and the corresponding single-nucleon cross section. These
``momentum distributions'' incorporate the role of FSI, and they can
be used to get the spectral function following the same procedure as
in the previous case (PWIA). Again, we perform a systematic analysis
by considering different values of Gaussian and Lorentzian widths. Finally,
the scaling function is directly evaluated from the spectral
function. We have compared these results with the one obtained
directly from the ratio between the inclusive $(e,e')$ cross section
(evaluated within the RMF+FSI model) and the single-nucleon cross
section taken at $p=|y|$ and ${\cal E}=0$. In all cases the results
obtained clearly show the essential role of FSI in producing the
required asymmetry in the scaling function, {\it i.e.,} to be
consistent with data. Concerning the second method we explored for
including FSI following \cite{33ankowski}, we note that its usage
leads only to qualitative description of the experimentally observed
asymmetry of the scaling function. As can be seen from
Figs.~\ref{fig09}--\ref{fig09b} the asymmetry due to the higher tail
of $f(\psi)$ for positive values of $\psi$ is not very different for
the three cases considered, the RFG model, the case with HO
single-particle wave functions and that with natural orbitals. In
addition, the obtained asymmetry is rather similar in the cases when
nonrelativistic or relativistic  optical potentials are used to
account for the FSI. Here we would like to emphasize that
considerably larger asymmetry is obtained in the RMF+FSI method (see
Figs.~\ref{fig10} and~\ref{fig11}), and especially when experimental
widths with the Lorentzian function are used in the calculations
(see Fig.~\ref{fig11}). So, in the case of the RMF+FSI method the
values of $f(\psi)$ for positive values of $\psi$ describe the
experimental data quite satisfactorily.

To summarize, our extensive study with different nuclear models
based on the independent particle shell model and even incorporating
short-range nucleon correlations through the use of NOs within the
Jastrow method, show different results for the momentum
distributions, particularly at high missing momenta where nucleon
correlations give rise to the presence of an important tail.
However, the scaling functions obtained from them do not present the
strong asymmetry at positive $\psi$ as given by the analysis of
data. We have proven that such a ``strong'' asymmetry in $f(\psi)$
does not emerge even when a broader energy-dependence is assumed in
the spectral function, but it does emerge when FSI are taken into
account. In particular, taking into account final-state interactions
in the RMF approach produces, in addition to a strong tail at
high-$p$ in the nucleon momentum distribution, a scaling function
with the right amount of asymmetry in accordance with data.

Here we would like to make some concluding remarks on the
results obtained in the present work. We established and considered
the relationship between the longitudinal scaling function $f(\psi)$
and the spectral function $S(p,{\cal E})$ accounting for the
restrictive condition of the PWIA in the description of the reaction
mechanism and, correspondingly, with the momentum distribution at
the specific conditions for the kinematically allowed region (the
excitation energy ${\cal E}_{\max} \rightarrow \infty$). We studied
the ingredients of the spectral function, namely the single-particle
momentum distributions $n_i(p)$ and the hole-state energy
distributions. In several consequent steps (starting with the IPSM)
with increasing complexity beyond the MFA we showed that the
$n_i(p)$ have to be considered within models in which the nucleon
correlations are taken into account. This leads to the existence of
high-momentum components of the momentum distributions. The
methodical study of the energy distribution of the hole states
showed that distributions with broader energy dependence (due to the
residual interaction) have to be considered, with single-particle
widths which are close to the empirically observed ones. The energy
distributions must have more complicated form than the Gaussian one,
e.g. the Lorentzian one. However, it was pointed out that all
mentioned conditions are clearly necessary, but are not sufficient
for a successful description of the experimentally observed
asymmetry of the scaling function $f(\psi)$, namely they give a
weaker asymmetry at positive values of $\psi$. We pointed out that
this characteristic feature of $f(\psi)$ can be reached only if the
FSI (and other peculiarities of the electron scattering beyond the
PWIA) are carefully taken into account. This was reached in the case
of RMF+FSI, while other methods using complex optical potentials for
FSI lead to results that have some discrepancies with the
experimental longitudinal scaling function $f(\psi)$ (e.g., for
$\psi < -0.9$ and $\psi > 1.2$), and they must be used with caution.
The obtained results open a new perspective on the extraction of
information on momentum distribution from experimental inclusive
$(e,e')$ data.

\subsection*{Acknowledgements}

This work was partially supported by DGI (MICINN-Spain) contracts
FIS2008-01301, FIS2008-04189, PCI2006-A7-0548, FPA2007-62216, FPA2010-17142,
 the Spanish Consolider-Ingenio programme CPAN (CSD2007-00042), by the
Junta de Andaluc\'{\i}a, and by the INFN-MICINN collaboration
agreements FPA2008-03770-E \& ACI2009-1053 ``Study of
relativistic dynamics in neutrino and electron scattering'', as well as by the
Bulgarian National Science Fund under Contract Nos. DO-02-285 and
DID--02/16--17.12.2009 and by Universidad Complutense de Madrid
(grupos UCM, 910059). One of the authors (M.V.I.) is grateful for
the warm hospitality given by the Universidad Complutense de Madrid
(UCM) and for support during his stay there from the State
Secretariat of Education and Universities of Spain (N/Ref.
SB2009-0007). This work is also supported in part (T.W.D.) by the
U.S. Department of Energy under cooperative agreement
DE-FC02-94ER40818.

\end{document}